\documentclass[9pt,twocolumn,twoside]{opticajnl}
\journal{opticajournal} 

\usepackage{caption}  
\usepackage{lineno}

\title{Manipulation of the orbital angular momentum of soft x-ray beams by consecutive diffractive optics}

\author[1,*]{NAZIR KHAN}
\author[2]{RAHUL JANGID}
\author[3]{TARAS STANISLAVCHUK}
\author[4]{AARON STEIN}
\author[2]{OLEG CHUBAR}
\author[2]{ANDI BARBOUR}
\author[3]{ANDREI SIRENKO}
\author[1]{VALERY KIRYUKHIN}
\author[2,$\dagger$]{CLAUDIO MAZZOLI}

\affil[1]{Department of Physics and Astronomy, Rutgers University, Piscataway, NJ 08854, USA }
\affil[2]{National Synchrotron Light Source II, Brookhaven National Laboratory, Upton, NY 11973, USA}
\affil[3]{Department of Physics, New Jersey Institute of Technology, Newark, NJ 07102, USA}
\affil[4]{Center for Functional Nanomaterials, Brookhaven National Laboratory, Upton, NY 11973, USA}
\affil[$\dagger$]{cmazzoli@bnl.gov}
\affil[*]{nazir.khan@rutgers.edu}

\begin{abstract}
Production and manipulation of orbital angular momentum (OAM) of coherent soft x-ray beams is demonstrated utilizing consecutive diffractive optics. OAM addition is observed upon passing the beam through consecutive fork gratings. The OAM of the beam was found to be decoupled from its spin angular momentum (SAM). Practical implementation of angular momentum control by consecutive devices in the x-ray regime opens new experimental opportunities, such as direct measurement of the beam’s OAM without resorting to phase sensitive techniques, including holography.  OAM analyzers utilizing fork gratings can be used to characterize the beams produced by synchrotron and free electron lasers sources; they can also be used in scattering experiments. 
\end{abstract}

\setboolean{displaycopyright}{false} 

\begin{document}

\maketitle

The angular momentum of light beam, which entails both spin and orbital components, plays an important role in the fundamental research \cite{1,2,3,4} as well as in the applied fields \cite{5,6,7,8,9,10} of optics. While the spin angular momentum $S_x$ is associated with the circular polarization states of light, the orbital angular momentum $L_x$ results from the azimuthal rotation of field’s phase around the beam axis ($x$) yielding a helical phase front. For general three-dimensional fields $S_x$ and $L_x$ are not separately independent and are always combined in the total angular momentum $J_x=S_x+L_x$ \cite{11,12,13}. However, in paraxial beams, both the polarization and phase rotation around the beam axis preserve the transversality of the electromagnetic field and give rise to valid independent angular momenta \cite{14,15}.  In the paraxial approximation, both \textit{S} and \textit{L} are intrinsic and longitudinal (i.e. parallel to the light propagation direction $x$) \cite{16}. The eigenstate of $S_x$ corresponds to uniform, circularly polarized beams where spin angular momentum (SAM) can take values of $\pm$ $\hbar$ per photon. The eigenstates of $L_x$ correspond to Laguerre-Gaussian (LG) modes of uniformly polarized beams having doughnut-like transverse intensity profile. Their dark centers coincide with the phase singularities of the electric field along the beam propagation direction, evolving with the azimuthal angle $\phi$ as exp(-i$\ell\phi$), where $\ell$ is an integer. For any given $\ell$, the beam has an orbital angular momentum (OAM) value of $\ell\hbar$ per photon.\\ 
\begin{figure*}[ht]
	\centering
	\includegraphics[scale=0.6]{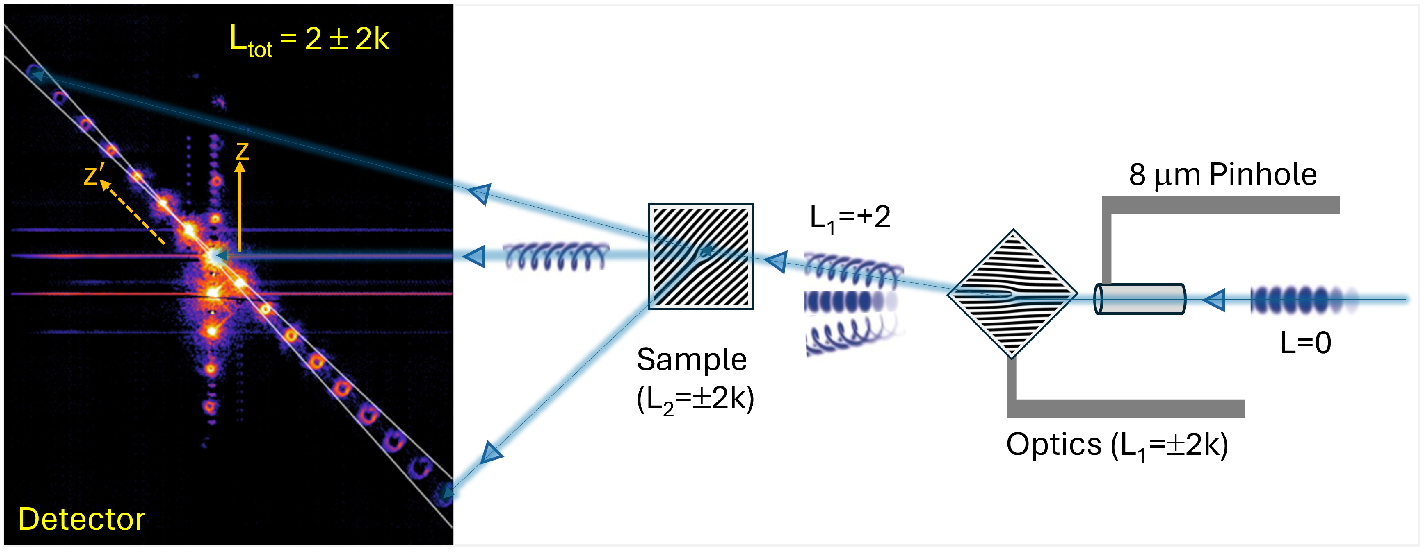}
	\caption{Schematic of the experimental set-up for the generation and manipulation of soft x-ray OAM beams using two consecutive diffraction fork gratings. The first grating ($L_1$) produces the OAM beams diffracted along the vertical $z$-direction, using the coherent gaussian beam from a synchrotron beamline collimated by an 8 $\mu$m pinhole.  One of these OAM beams, $L_1$=+2$\hbar$, is then selected to impinge on the second fork grating ($L_2$) used as the sample which diffracts the beams with modified OAM along the $z'$-direction.  Both diffraction patterns are observed on the area detector in straight-through configuration.  In the experiment shown, the two-step diffraction setup yields the beams with total OAM given by $L_{\rm tot}$=(2$\pm$2k)$\hbar$, where $k$=0, 1, 2, 3, etc. Intersecting white lines on the detector image are added to illustrate the increasing size of the diffracted beams with increasing diffraction order away from the $L_{\rm tot}$ =0 position.\label{fig1}}
\end{figure*}
\indent SAM of light finds many applications. In the x-ray regime, a well-known example is x-ray magnetic circular dichroism utilized for studies of magnetic materials. Practical and scientific applications of OAM are less explored. This subject is currently attracting significant attention. LG beams have been generated in spectral ranges from THz \cite{17,18}  to the shorter wavelengths of extreme ultraviolet \cite{19,20,21,22,23}, soft x-ray \cite{24,25}, and hard x-ray \cite{26} radiation. Soft x-rays are capable of probing electronic states of relevant ions in matter via resonant absorption and scattering, in addition to topological states of matter such as stripes observed in high-temperature superconductors, skyrmions and vortices in magnets, and vortex polarization domains in ferroelectrics \cite{27,28}. Introduction and control of OAM in synchrotron-generated coherent soft x-rays with high brilliance could potentially enable the development of new advanced OAM-based spectroscopic and scattering techniques. In this work, we demonstrate that the LG modes of synchrotron-based coherent soft x-rays can be created and manipulated by consecutive diffractive fork gratings, which is a necessary prerequisite for advanced applications of OAM in the x-ray regime.\\
\indent The experiment was performed at the 23-ID-1 (CSX) beamline at the NSLS-II (National Synchrotron Light Source II) \cite{29,30}. The beamline provides ultrabright coherent x-rays with energies between 0.25 keV and 2 keV, produced by two identical elliptical polarized undulator sources that allow for complete polarization control, enabling any combination of linear and circular polarization.  A set of pinholes inside and outside of the TARDIS scattering chamber grant the extraction of the coherent fraction of the x-ray beam. Due to the brilliance of the NSLS-II source, typically a photon flux of 10$^{13}$ ph/s impinges on the sample with a beam footprint of ten microns. The high degree of transverse coherence obtained on the CSX beamline is the key for the OAM beam generation. The schematic of the transmission experimental setup inside the TARDIS scattering chamber is shown in Fig. \ref{fig1}. We started with a gaussian beam ($L$=0) of x-ray of energy 706 eV set by the beamline’s monochromator. The beam is then allowed to pass through an 8 $\mu$m pinhole to produce a paraxial incident x-ray beam with high spatial coherence impinging on the first diffractive optical element labeled as optics in Fig. \ref{fig1}. The shown optical element is a 2-fork dislocation grating ($L_1$=$\pm$2k) which is one of the 15 diffractive fork gratings contained in a chip called OAM optics [see Fig. S1(a) and the text in the supplemental material]. By translating the OAM optics chip across the beam from the pinhole, a different fork grating can be selected as optics. The pinhole size is chosen to homogenously illuminate the optics by a coherent x-ray beam, which is necessary to provide a defined OAM charge to the beam. The optics generates a number of LG modes in their scattering plane (see $z$ in Fig. \ref{fig1}), each of them carrying different OAM charges. For every experiment, one of these LG modes is selected to impinge on a second diffractive fork grating, labeled as sample in Fig. \ref{fig1}. This chosen sample grating is one of the 9 diffractive fork gratings contained in a chip called OAM sample [see Fig. S1(b) and the text in the supplemental material].  By translating the OAM sample chip across the impinging beam, a different fork grating can be used as sample to further manipulate the OAM beam properties. The optics and the sample are placed 8.65 mm and 14.19 mm downstream of the 8 $\mu$m pinhole, respectively. A FastCCD area detector located 350 mm downstream of the sample was used to capture the diffraction patterns spreading along $z$ and $z'$ directions generated by the optics and the sample, respectively. These directions differ by 45 degrees to minimize the unwanted overlap of the diffraction patterns coming from the different gratings. In the cartoon experiment shown in Fig. \ref{fig1}, the 2-fork dislocation grating generates first order diffracted LG modes with OAM charges $L_1$=$\pm$2$\hbar$. The $L_1$=+2$\hbar$ LG mode is then used to illuminate a 2-fork grating as the sample.  Then OAM addition rules stipulate that the LG modes generated by the sample should carry OAM charges given by $L_{\rm tot}$ = (2$\pm$2k)$\hbar$, where $k$ is an integer. LG modes with OAM charge as large as $L_{\rm tot}$ =20$\hbar$ can be seen on the detector. Though the experiment was performed at x-ray energy of 706 eV (Fe $L_3$ edge); these gratings can be used in the energy range from $\sim$500 eV (oxygen K-edge) to $\sim$1 keV (covering all the Transition Metal 3d $L$ edges relevant in quantum magnetism).\\
\indent The diffraction pattern from a single fork grating exhibits a Gaussian mode for the transmitted beam where $k$=0, and doughnut-like intensity profiles for the diffracted beams where, $k\neq$0. The size of the doughnuts linearly increases with $|k|$. This is clearly observed for the $z'$ axis data in Fig. \ref{fig1}. Importantly, the Gaussian beam is observed in a non-zero diffraction order of the second grating. The diffraction pattern along $z'$ is therefore shifted from the position of the impingent beam. As described below, this is direct evidence of the OAM addition in the two-grating setup. This shows that the quantized OAM of the photons remains well defined after multiple diffraction events, and that the OAM charge of the various beams generated through the consecutive fork gratings changes according to the standard algebraic addition rules in the process. Thus, our measurements show that the OAM of x-ray beams can be produced and manipulated by consecutive optical elements.\\
\indent To demonstrate the angular momentum addition for various OAM charges, we did measurements using different combinations of fork gratings. Additionally, different incident beam polarizations (linear, left and right circular) were tested. Below, we show quantitative analysis of the obtained diffraction patterns along the $z'$ axis. Figure \ref{fig2}(a) exhibits detector images of the diffraction patterns (two top panels) generated by a 1-fork dislocation grating illuminated by LG beams carrying OAM charges $L_1$=$\pm$4$\hbar$ and diffraction patterns (three bottom panels) generated by a 2-fork grating illuminated by LG beams carrying OAM charges $L_1$= -1$\hbar$, 2$\hbar$, and 3$\hbar$. The polarization state of the soft x-ray beams from the undulator was set to right circular polarization (C+). A clear asymmetry on the two sides of the most intense central transmitted beam (dashed vertical line) illuminating the sample can be seen in all the diffraction patterns, showing that the beam impinging on the sample carries a nonzero OAM charge. Furthermore, this additional OAM charge propagates in the diffracted orders of the sample grating ($z'$),  following the algebraic addition of OAM charges. By tracking the shift and indexing the OAM charges of the diffracted beams from the second grating, OAM addition in the second diffraction event can be confirmed. This is demonstrated by the data in Figs. \ref{fig2}(b) and \ref{fig2}(c) showing the intensity profiles along the $z'$ axis for the linecut through the center of the doughnuts in the diffraction patterns generated by a 1-fork grating illuminated by the beam of charge $L_1$=+4$\hbar$ and -4$\hbar$, respectively. The double-peak intensity profile of a diffraction order indicates an LG mode carrying OAM, whereas the symmetric narrow single peak indicates the Gaussian beam with zero OAM. A Gaussian mode (index $\ell$=0) appears at the 4$^{th}$ diffraction order on the left (right) side of the central beam of the diffraction pattern when the impinging beam carries the OAM charge $L_1$=+4$\hbar$ ($L_1$= -4$\hbar$), see the shaded peaks in Figs. \ref{fig2}(b) and \ref{fig2}(c), respectively.  This proves the OAM addition at the second diffraction event. Note, the peaks in the first two diffraction orders are unresolved, but their widths, as well as the peak separations in the resolved peaks, allow unambiguous identification of the $\ell$=0 position.\\
\begin{figure}[t]
\centering
\includegraphics[width=\linewidth]{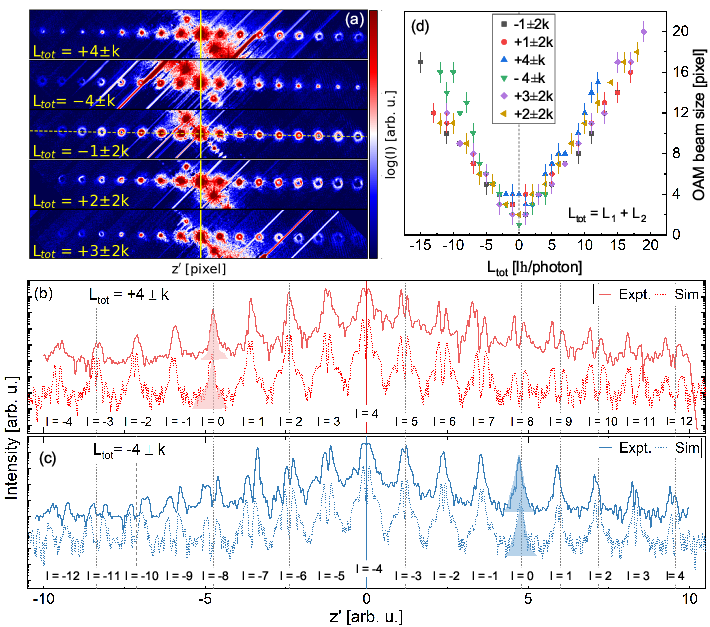}
\caption{(a) Detector images of the diffraction patterns containing the OAM beams generated by different combinations of consecutive fork gratings. OAM beams of different momenta ($L_1$= -4$\hbar$, 4$\hbar$, -1$\hbar$, 2$\hbar$, 3$\hbar$) are used selectively to illuminate the 2nd grating. The diffraction from the 2nd grating changes the OAM of the incident beam according to the OAM addition rule. (b) and (c) Solid lines depict the intensity profiles of the linecuts through the centers of the OAM beams generated by 1-fork grating illuminated by OAM beams with $L_1$=4$\hbar$ and -4$\hbar$, respectively. Different LG modes of the diffracted beams are indexed by the orbital angular momentum quantum number $\ell$. Gaussian beams ($\ell$=0) are shaded. Vertical solid lines indicate the positions of the transmitted beams from the second grating at $z'$=0. Dashed lines show the numerical simulations, as described in the text. (d) The OAM beam size as function of the total OAM, $L_{\rm tot}$=$L_1$+$L_2$. Vertical error bars reflect experimental resolution.\label{fig2}}
\end{figure}
\indent The peak separation within each LG mode (and, correspondingly, the beam image size on the detector) increases with the distance from the $\ell$=0 peak. Thus, the diffracted beam size is related to its OAM charge. Fig. \ref{fig2}(d) shows the beam size as a function of its total orbital angular momentum $L_{\rm tot}$=$L_1$+$L_2$ for multiple combinations of the first and second fork gratings and, therefore, for multiple combinations of $L_1$ and $L_2$. A uniform error bar of 2 detector pixels, which is roughly the FWHM of the gaussian peaks (index $\ell$=0) in Figs. \ref{fig2}(b, c), is assigned to the OAM beam size. A linear dependence of the OAM on the beam size is found. Moreover, the data from different experimental setups fall on the same lines, within the measurement error.  (Small deviations in higher diffraction orders for $L_{\rm tot}$=($\pm$4$\pm$k)$\hbar$ could be attributed to misalignment of the $L_1$=$\pm$4$\hbar$ OAM beams with the 1-fork second grating due to higher divergence of the $L_1$=$\pm$4$\hbar$ beams.) Thus, the size of the diffracted beam provides direct measure of $L_{tot}$  in our experiments. Numerical simulations \cite{31,32} confirm the given interpretation of experimental results. Diffraction intensities were calculated for the different combinations of the fork gratings used in our experiments [Figs. S2(a-c)]. Intensity profiles of the linecuts through the centers of the doughnut-like LG modes [Figs. S2(b, c)] were compared with the experimental data and excellent agreement was found. Simulation results for $L_1$=+4$\hbar$ and $L_1$=-4$\hbar$ beams impingent on a 1-fork grating are shown in Figs. \ref{fig2}(b, c). The calculated and experimental peak shapes are in good agreement, and the calculated diffraction order indexes (and the corresponding OAM) coincide with the experimental values.\\
\begin{figure}[t]
\centering
\includegraphics[width=\linewidth]{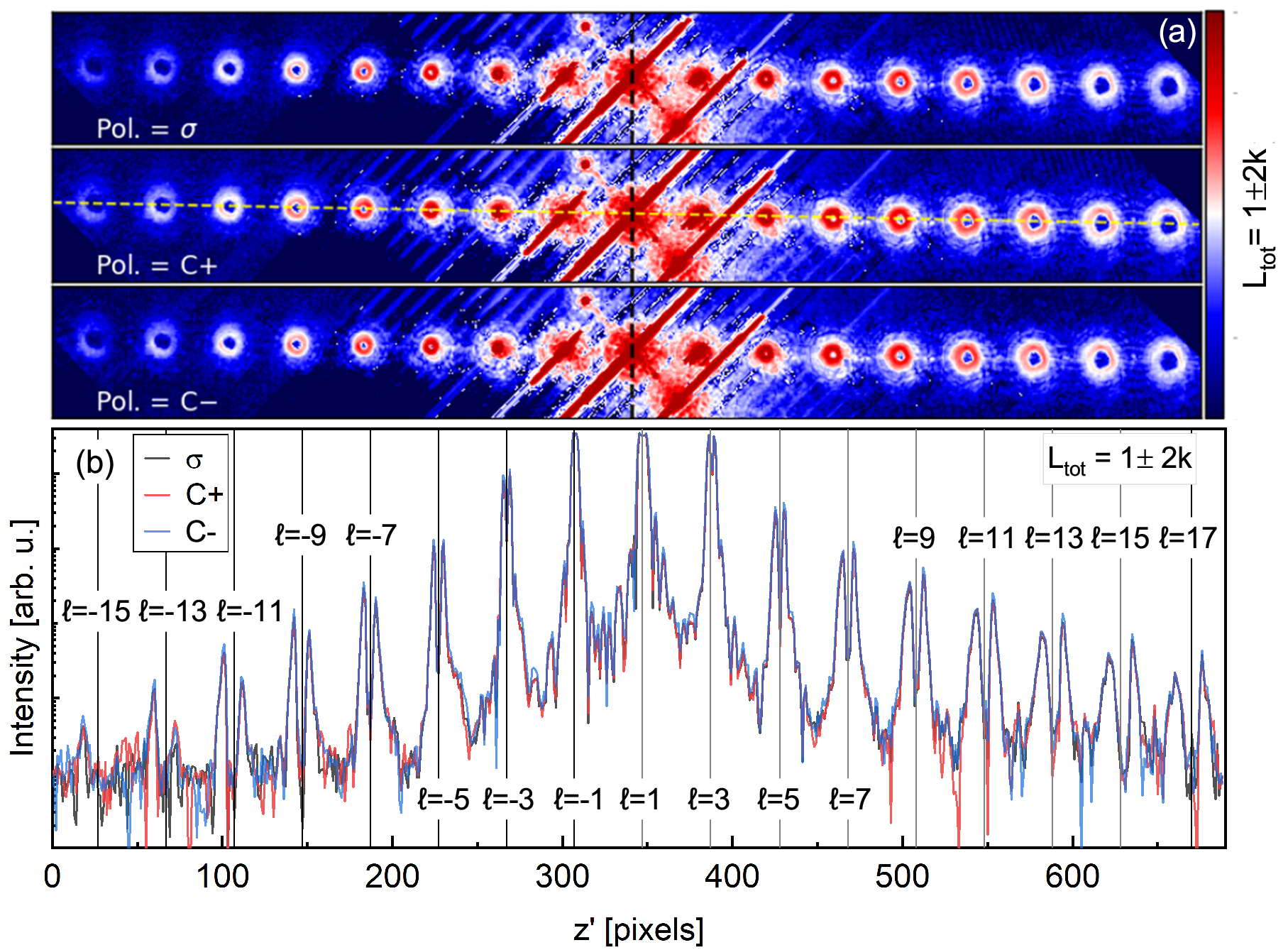}
\caption{(a) Detector images of the diffraction patterns produced by a 2-fork grating illuminated by $\ell$=+1 beam generated by 1-fork grating. The three panels correspond to different polarization states of the incoming beam from the synchrotron beamline, as indicated.  (b) Line profiles through the centers of the diffracted beams shown in (a). \label{fig3}}
\end{figure}
\indent The total angular momentum of light entails two components, SAM and OAM; they are uncoupled in paraxial LG beams. To check that OAM characteristics are independent of the polarization states of the soft x-ray beam and the paraxial approximation works well for our transmission setup, we have used different polarization states of the soft x-ray generated by the undulator in our experiments: the horizontal linear ($\sigma$), right (C+) and left (C-) circular polarizations. We find that the polarization state has no effect on the OAM-related features. Figure \ref{fig3} shows an example experiment utilizing consecutive 1-fork and 2-fork gratings producing the total OAM $L_{\rm tot}$=(1$\pm$2k)$\hbar$. The intensity profiles of the linecuts through the center of the doughnuts for the three different polarization states are practically identical. Thus, the OAM and SAM are indeed uncoupled and can be controlled separately.\\
\indent LG beams were previously produced in the soft x-ray regime by diffractive optics \cite{24}. In our work, this approach is extended to consecutive optical devices. It is found that the x-ray photons preserve their quantum characteristics after several interactions with diffractive optics. This allows significant flexibility in manipulating the beam’s OAM. Multiple units of $\hbar$ can be added to or subtracted from the OAM of the x-ray photons. Large OAM charges can be therefore created, if necessary. This is useful because large-k fork gratings and spiral zone plates producing high OAM require large transverse coherence (all the dislocation-type defects must be illuminated coherently and homogeneously), which may be difficult to achieve. Multiple in-line devices enable useful practical applications. For example, OAM beam can be created first, let interact with an object (sample), and then be analyzed by another device. This opens new possibilities for studies of interactions of OAM beams with matter, similar to what is routinely done now using SAM. In particular, our experiments show that fork gratings can be used as analyzers providing direct determination of the beam’s OAM charge by observation of the characteristic displaced diffraction patterns. Importantly, this is done without using complex setups based on holography, typically required to analyze phase-dependent phenomena. Our measurements confirm decoupling of OAM and SAM in the beams manipulated by fork gratings. Consecutive modification of OAM and SAM is known to produce beams carrying fractional (half-integer) total angular momentum via nonlinear upconversion \cite{33,34}. This has been demonstrated in crossed beam geometries. Our work opens possibilities to apply this approach utilizing in-line optics. Our approach should be extendable to various optical elements, such as spiral zone plates, phase retarders, etc. Such elements should be immediately useful for characterization of the beams produced by helical undulators \cite{19}, and for free electron lasers \cite{35,36}. We also envision that simple grating-based OAM analyzers will find applications in novel spectroscopic techniques for studies of quantum magnetism, topological matter, as well as magnetic and structural mesoscopic chirality/vorticity \cite{25,27,28,37,38}.  

\begin{backmatter}
\bmsection{Funding} This work was supported by the DOE under Grant No. DE-FG02-07ER46382.

\bmsection{Acknowledgment} VK and AS were supported by Keck Foundation. This research used resources of the National Synchrotron Light Source II, a DOE Office of Science User Facility operated by Brookhaven National Laboratory under Contract No. DE-SC0012704 and resources made available through BNL/LDRD\# 19-013.

\bmsection{Disclosures} The authors declare no conflicts of interest.

\bmsection{Data availability} Data underlying the results presented in this paper may be obtained from the authors upon reasonable request.

\bmsection{Supplemental document}
See Supplement 1 for supporting content.
\end{backmatter}



\begin{thebibliography}{1}
\bibitem{1}	A. Mair, A. Vaziri, G. Weihs, \textit{et al}., \href{https://www.nature.com/articles/35085529}{Nature} \textbf{412}, 313–316 (2001).
\bibitem{2}	G. Molina-Terriza, J. P. Torres, and L. Torner, \href{https://www.nature.com/articles/nphys607}{Nat. Phys.} \textbf{3}, 305–310 (2007).
\bibitem{3}	J. Leach, B. Jack, J. Romero, \textit{et al}., \href{https://www.science.org/doi/10.1126/science.1190523}{Science} \textbf{329}, 662–665 (2010).
\bibitem{4}	Y. Shen, X. Wang, Z. Xie, \textit{et al}., \href{https://www.nature.com/articles/s41377-019-0194-2}{Light Sci. Appl.} \textbf{8}, 90 (2019).
\bibitem{5}	B. Thid\'{e}, H. Then, J. Sj\"{o}holm, \textit{et al}., \href{https://journals.aps.org/prl/abstract/10.1103/PhysRevLett.99.087701}{Phys. Rev. Lett.} \textbf{99}, 087701 (2007).
\bibitem{6}	J. Romero, D. Giovannini, S. Franke-Arnold, \textit{et al}., \href{https://doi.org/10.1103/PhysRevA.86.012334}{Phys. Rev. A} \textbf{86}, 012334 (2012).
\bibitem{7}	N. Uribe-Patarroyo, A. Fraine, D. S. Simon, \textit{et al}., \href{https://doi.org/10.1103/PhysRevLett.110.043601}{Phys. Rev. Lett.} \textbf{110}, 043601 (2013).
\bibitem{8}	J. T. Barreiro, T.-C. Wei, and P. G. Kwiat, \href{https://www.nature.com/articles/nphys919}{Nat. Phys.} \textbf{4}, 282–286 (2008).
\bibitem{9}	J. B\'{e}gin, A. Jain, A. Parks, \textit{et al}., \href{https://www.nature.com/articles/s41566-022-01100-0}{Nat. Photonics.} \textbf{17}, 82 (2023).
\bibitem{10} Y. Shen, Q. Zhan, L. G Wright, \textit{et al}., \href{https://iopscience.iop.org/article/10.1088/2040-8986/ace4dc}{J. Opt.} \textbf{25}, 093001 (2023).
\bibitem{11} S. J. van Enk and G. Nienhuis, \href{https://iopscience.iop.org/article/10.1209/0295-5075/25/7/004}{Europhys. Lett.} \textbf{25}, 497–501 (1994).
\bibitem{12} C. Cohen-Tannoudji, J. Dupont-Roc, and G. Grynberg, \textit{Photons and Atoms: Introduction to Quantum Electrodynamics} (Wiley, New York, 1997), 486 pp.
\bibitem{13} L. Allen, S. Barnett, and M. Padgett, Eds., \textit{Optical Angular Momentum} (Institute of Physics, Bristol, UK, 2003).
\bibitem{14} L. Allen, M. W. Beijersbergen, R. J. C. Spreeuw, \textit{et al}., \href{https://doi.org/10.1103/PhysRevA.45.8185}{Phys. Rev. A} \textbf{45}, 8185–8189 (1992).
\bibitem{15} S. M. Barnett, \href{https://iopscience.iop.org/article/10.1088/1464-4266/4/2/361}{J. Opt. B: Quantum Semiclass. Opt.} \textbf{4}, S7–S16 (2002).
\bibitem{16} K. Y. Bliokh and F. Nori, \href{https://doi.org/10.1016/j.physrep.2015.06.003}{Phys. Rep.} \textbf{592}, 1 (2015).
\bibitem{17} A. A. Sirenko, P. Marsik, C. Bernhard, \textit{et al}., \href{https://doi.org/10.1103/PhysRevLett.122.237401}{Phys. Rev. Lett.} \textbf{122}, 237401(2019).
\bibitem{18} A. A. Sirenko, P. Marsik, L. Bugnon, \textit{et al}., \href{https://doi.org/10.1103/PhysRevLett.126.157401}{Phys. Rev. Lett.} \textbf{126}, 157401 (2021).
\bibitem{19} J. Bahrdt, K. Holldack, P. Kuske, \textit{et al}., \href{https://doi.org/10.1103/PhysRevLett.111.034801}{Phys. Rev. Lett.} \textbf{111}, 034801 (2013).
\bibitem{20} G. Gariepy, J. Leach, K. Taec Kim, \textit{et al}., \href{https://doi.org/10.1103/PhysRevLett.113.153901}{Phys. Rev. Lett.} \textbf{113}, 153901 (2014).
\bibitem{21} R G\'{e}neaux, A. Camper, T. Auguste, \textit{et al}., \href{https://www.nature.com/articles/ncomms12583}{Nat. Commun.} \textbf{7}, 12583 (2016).
\bibitem{22} P. R. Ribi\v{c}, B. R\"{o}sner, D. Gauthier, \textit{et al}., \href{https://doi.org/10.1103/PhysRevX.7.031036}{Phys. Rev. X}. \textbf{7}, 031036 (2017).
\bibitem{23} L. Rego, K. M. Dorney, N. J. Brooks, \textit{et al}., \href{https://www.science.org/doi/10.1126/science.aaw9486}{Science}, \textbf{364}, eaaw9486 (2019).
\bibitem{24} J. C. T. Lee, S. J. Alexander, S. D. Kevan, \textit{et al}., \href{https://www.nature.com/articles/s41566-018-0328-8}{Nat. Photonics} \textbf{13}, 205 (2019).
\bibitem{25} J. S. Woods, X. M. Chen, R. V. Chopdekar, \textit{et al}., \href{https://doi.org/10.1103/PhysRevLett.126.117201}{Phys. Rev. Lett.} \textbf{126}, 117201 (2021).
\bibitem{26} A. G. Peele, K. A. Nugent, A. P. Mancuso, \textit{et al}., \href{https://opg.optica.org/josaa/fulltext.cfm?uri=josaa-21-8-1575&id=80716}{J. Opt. Soc. Am. A} \textbf{21}, 1575 (2004).
\bibitem{27} M. van Veenendaal, \href{https://doi.org/10.1103/PhysRevB.92.245116}{Phys. Rev. B} \textbf{92}, 245116 (2015).
\bibitem{28} P. Gao, J. Britson, C. T. Nelson, \textit{et al}., \href{https://www.nature.com/articles/ncomms4801}{Nat. Commun.} \textbf{5}, 3801 (2014).
\bibitem{29} D. Allan, T. Caswell, S. Campbell, \textit{et al}., \href{https://doi.org/10.1080/08940886.2019.1608121}{Sync. Rad. News} \textbf{32}, 19 (2019).
\bibitem{30} https://github.com/NSLS-II-CSX/csxtools
\bibitem{31} O. Chubar, G. Williams, Y. Gao, \textit{et al}., \href{https://doi.org/10.1364/JOSAA.473367}{J. Opt. Soc. Am. A} \textbf{39}, C240-C252 (2022).
\bibitem{32} \href{https://github.com/ochubar/SRW}{SRW GitHub repository}
\bibitem{33} K. E. Ballantine, J. F. Donegan, and P. R. Eastham, \href{https://www.science.org/doi/10.1126/sciadv.1501748}{Sci. Adv.} \textbf{2}, e1501748 (2016).
\bibitem{34} M. Luttmann, M. Vimal, M. Guer, \textit{et al}., \href{https://www.science.org/doi/10.1126/sciadv.adf3486}{Sci. Adv.} \textbf{9}, eadf3486 (2023).
\bibitem{35} E. Hemsing, P. Musumeci, S. Reiche, \textit{et al}., \href{https://doi.org/10.1103/PhysRevLett.102.174801}{Phys. Rev. Lett.} \textbf{102}, 174801 (2009).
\bibitem{36} E. Hemsing, A. Knyazik, M. Dunning, \textit{et al}., \href{https://www.nature.com/articles/nphys2712}{Nat. Phys.} \textbf{9}, 549–553 (2013).
\bibitem{37} M. Fanciulli, M. Pancaldi, E. Pedersoli, \textit{et al}., \href{https://doi.org/10.1103/PhysRevLett.128.077401}{Phys. Rev. Lett.} \textbf{128}, 077401 (2022).
\bibitem{38} M. Fanciulli, M. Pancaldi, A. Stanciu, \textit{et al}., \href{https://journals.aps.org/prl/abstract/10.1103/PhysRevLett.134.156701}{Phys. Rev. Lett.} \textbf{134}, 156701 (2025).
  
\end{thebibliography}
\end{document}